# Quantifying the Impact on Software Complexity of Composable Inductive Programming using Zoea

Edward McDaid FBCS
Chief Technology Officer
Zoea Ltd.

Sarah McDaid PhD
Visiting Senior Research Fellow
London South Bank University

**Abstract**

Composable inductive programming as implemented in the Zoea programming language is a simple declarative approach to software development. At the language level it is evident that Zoea is significantly simpler than all mainstream languages. However, until now we have only had anecdotal evidence that software produced with Zoea is also simpler than equivalent software produced with conventional languages. This paper presents the results of a quantitative comparison of the software complexity of equivalent code implemented in Zoea and also in a conventional programming language. The study uses a varied set of programming tasks from a popular programming language chrestomathy. Results are presented for relative program complexity using two established metrics and also for relative program size. It was found that Zoea programs are approximately 50% the complexity of equivalent programs in a conventional language and on average equal in size. The results suggest that current programming languages (as opposed to software requirements) are the largest contributor to software complexity and that significant complexity could be avoided through an inductive programming approach.

## 1. Introduction

Inductive programming and related fields have been the subject of computer science research for over five decades [1,2,3,4]. The key concept in inductive programming is that programs are specified using examples of inputs and outputs rather than by describing the algorithm steps with a conventional programming language. The compiler for an inductive programming language generates the solution from the input-output cases rather than effectively translating the source language into a different target language via a parse tree.

Until recently inductive programming has been considered appropriate only for the generation of small programs. Composable inductive programming as introduced with the Zoea language extends inductive programming in a number of ways [5]. These include the ability to compose smaller programs to form larger ones as well as the ability to specify intermediate values between inputs and outputs. These together with other improvements allow for the creation of software of any size.

Listing 1 shows an example of a program in Zoea which determines whether the input is a palindrome. Here we can see that the program contains no instructions for how to ascertain whether a string is a palindrome. Rather it is structured like a set of test cases with associated inputs and outputs. The Zoea compiler uses a combination of programming knowledge and pattern matching to determine the required code from the test cases.

```
program: palindrome
  case: 1
    input: abcdcba
    output: true
  case: 2
    input: x
    output: true
  case: 3
    input: dog
    output: false
  case: 4
    input: ''
    output: false
```

**Listing 1**

## 2. Software complexity

Software complexity is an aspect of software development that is encountered by most software developers at some stage. The idea that some programs are more complicated than others is intuitive in the same way that some programs are larger than others. It is widely accepted that more complicated software takes more effort to develop, has more defects and costs more to maintain than does simpler software. Software complexity has been the subject of much academic research and a number of software complexity metrics have been developed. This reflects the fact that software can be complex in a number of different ways.

McCabe's cyclomatic complexity considers the structural complexity of a program [6]. This is calculated for each software unit such as a program or subroutine by reducing the control flow to a skeleton and then counting the number of distinct control flow paths. Every linear sequence of program statements gives rise to a single flow. Different conditional and control flow structures give rise to additional numbers of flows. For example an if-then-else construct adds two additional flows on top of the main flow within which the statement is embedded. The complexity of a complete program is simply the sum of the complexities for all constituent units. The resulting figure corresponds to the number of distinct paths that can be traced through the code at runtime.

Halstead difficulty is another software complexity measure [7]. Unlike cyclomatic complexity this does not consider the structure of the program but rather is formulated in terms of operators and operands. Operators are tokens that are part of the programming language while operands are variables, constants or values that are specific to a given problem. Difficulty is defined as:

$$D = (n1 / 2) + (n2 / N2)$$

where n1 is the number of distinct operators, n2 is the number of distinct operands and N2 is the total number of operands. The motivation of this approach was to support different languages while avoiding the specifics of individual platforms.

Research has also questioned and explored the nature of software complexity. Brooks [8] makes a distinction between accidental and essential complexity. Essential complexity is regarded as inherent in a given problem and cannot be removed. Accidental complexity is not an intrinsic part of a problem but rather is introduced by people, tools or technology related to solving the problem. Brooks also expresses the view that accidental complexity has become less of an overhead in modern software development with most complexity now encountered being essential.

**3. Programming language complexity**

The question of whether a given programming language is more or less complex in some sense than another programming language is a frequent topic of discussion for software developers. Surprisingly this question seems to have been the subject of relatively little serious scientific inquiry.

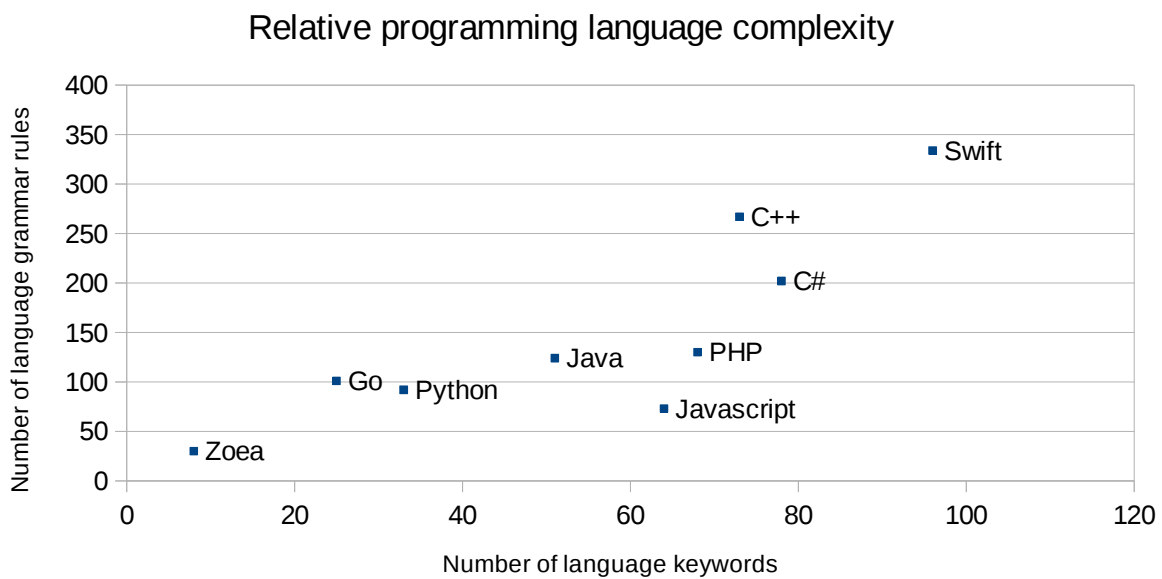

**Figure 1**

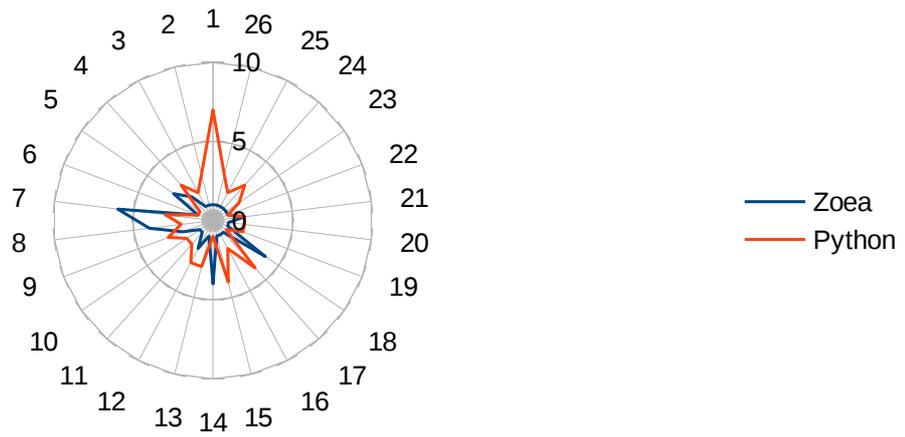

**Figure 2 – Cyclomatic Complexity (radial) By Task (angular)**

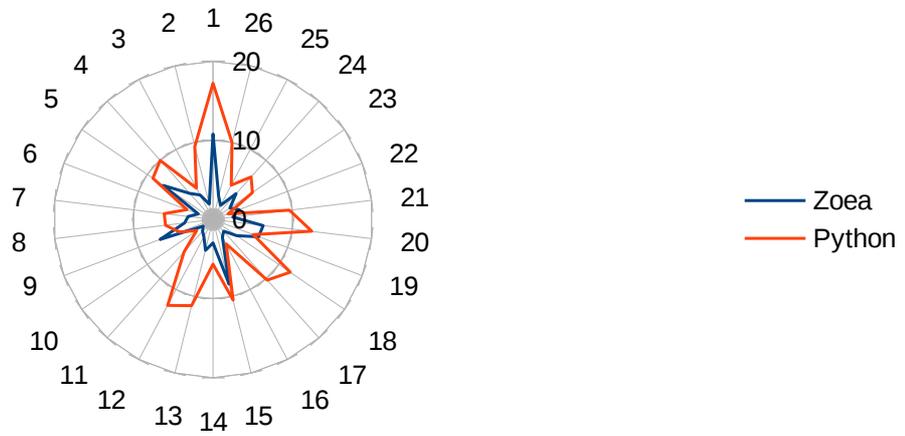

**Figure 3 – Halstead Difficulty (radial) By Task (angular)**

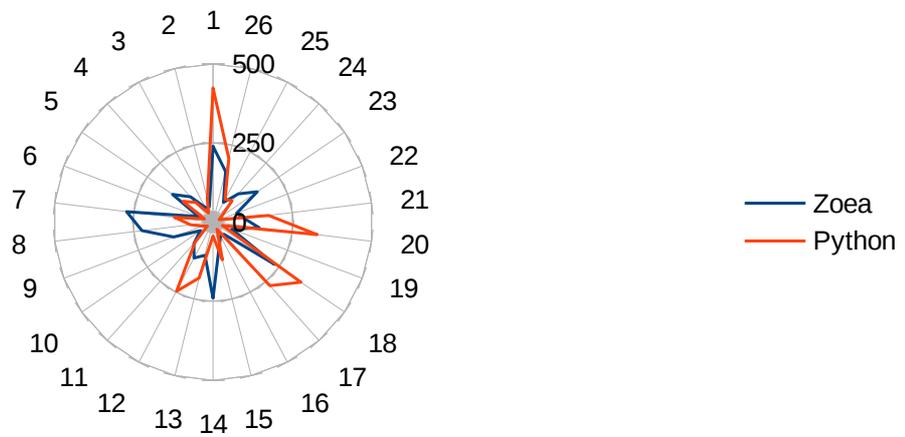

**Figure 4 – Program Size In Bytes (radial) By Task (angular)**

It is clear that any direct metrics relating to programming languages would of necessity have to relate to the intrinsic properties of languages such as the syntax or semantics. MacLennan [9] proposed a number of syntactic metrics for programming languages based on the size of programming language grammars as well as other metrics relating to semantics and transformation. Singer et al. [10] additionally used the number of language reserved words (or keywords) as a measure of complexity as well as the relative size of programming language manuals. Vanderwiel et al. [11] suggested using the complexity of equivalent programs in different languages as an indirect means of assessing the relative complexity of the implementation languages.

Figure 1 shows the relative programming language complexity of Zoea compared with a number of popular conventional programming languages using a combination of two language metrics. For each programming language this plots the number of rules in the language grammar against the number of reserved words in the language. The use of two metrics in this way serves to reduce any bias associated with a single metric and also to validate the result for each metric against the other. This shows that Zoea has around 25% the number of keywords and 30% the number of grammar rules compared with the simplest conventional languages. It can also be seen that the two metrics are in broad agreement in each case.

## 4. Study

The primary objective of the study is to answer the following question. Are programs written using Zoea more or less complex than equivalent programs written in a conventional programming language?

In order to reduce bias the study utilises multiple tasks with each task implemented both in Zoea and a conventional programming language. In order to maximise the number of tasks that could be included it was decided to use a single conventional language. Python was selected as it is both popular and relatively simple when compared with other conventional languages.

The tasks together with the corresponding Python and Zoea code were obtained from Rosetta Code [12] which is a popular programming language chrestomathy. Rosetta Code currently (May 2020) provides examples of 1017 tasks with each task typically implemented in multiple programming languages. Across all of the tasks, solutions are provided in 785 different programming languages. Some languages such as Python are represented in solutions to many tasks. Rosetta Code has been used in previous comparative studies of programming languages [13].

The 26 tasks included in the study were all of those for which both Python and Zoea solutions were available. Some tasks have multiple solutions in a given language in which case the simplest solution was selected.

A number of open source tools for measuring the Halstead difficulty of Python programs were evaluated. However, manual verification of their output showed that all of these produced inaccurate results. Instead a simple custom tool was created for this purpose. The results from this tool were also verified manually.

Cyclomatic complexity of Python programs was measured using 'multimetric 1.1.14' [14]. Zoea includes tools that measure cyclomatic complexity and Halstead difficulty. For both languages comments were removed before the program size was measured.

## 5. Results

Figure 2 shows the results for cyclomatic complexity for the Python and Zoea code for each of the tasks. Python is shown as a red line and Zoea as a blue line. Figure 3 shows a similar plot for Halstead difficulty. In both of these charts it is clear that the Zoea values are generally less than the Python values although the cyclomatic complexity plot has more outliers. The main reason for this is that some of the Zoea programs were over-specified in terms of the number of cases that they contain. It would have been possible for these tasks to reduce the number of cases in the Zoea programs to make them smaller and simpler while retaining the same behaviour but this action was not taken.

Figure 4 shows a plot of the program sizes for the tasks in bytes. Any trend in this case is less obvious visually.

Figures 5-7 present plots of the Python/Zoea ratios for each task for cyclomatic complexity, Halstead difficulty and program size respectively. Note that the tasks have been reordered so that the ratio values are in ascending order in each plot.

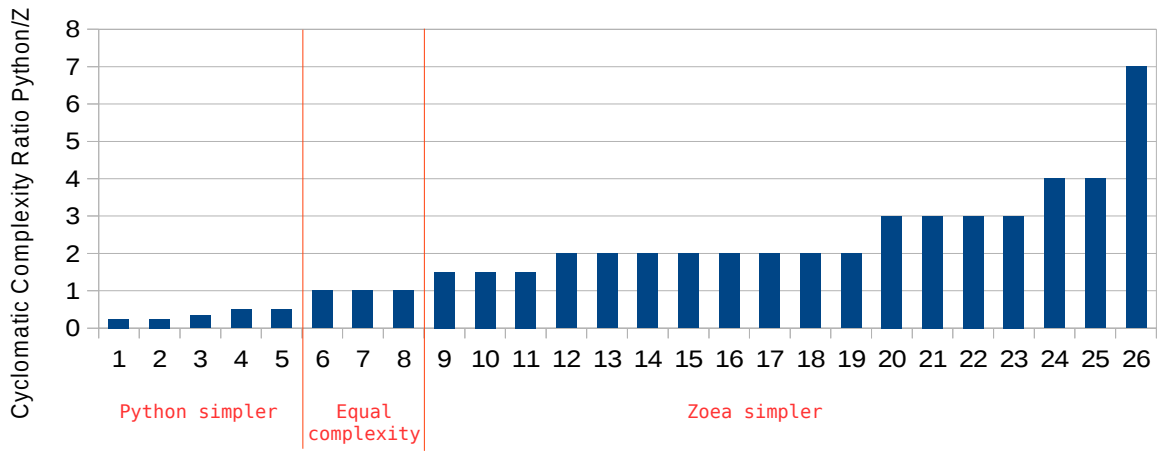

**Figure 5 – Cyclomatic Complexity Ratio Python/Zoea Ranked By Task**

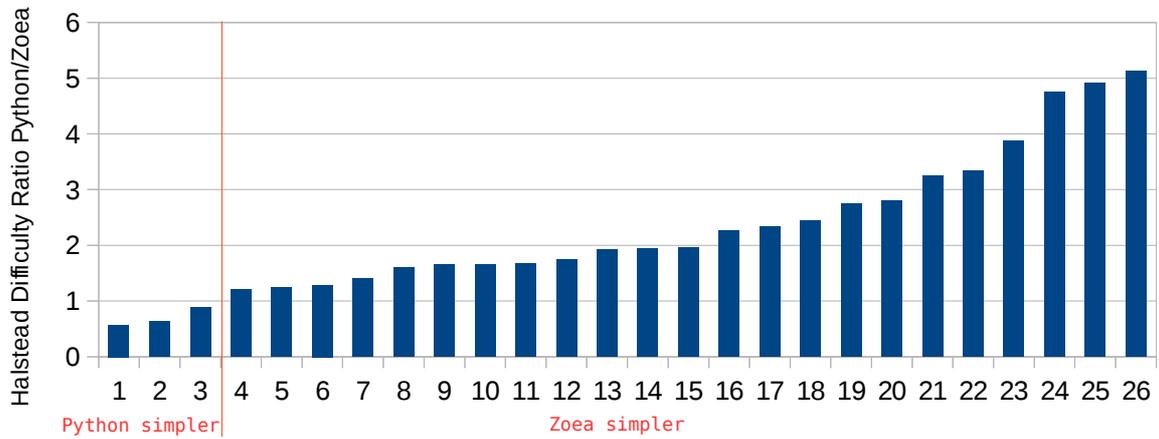

**Figure 6 – Halstead Difficulty Ratio Python/Zoea Ranked By Task**

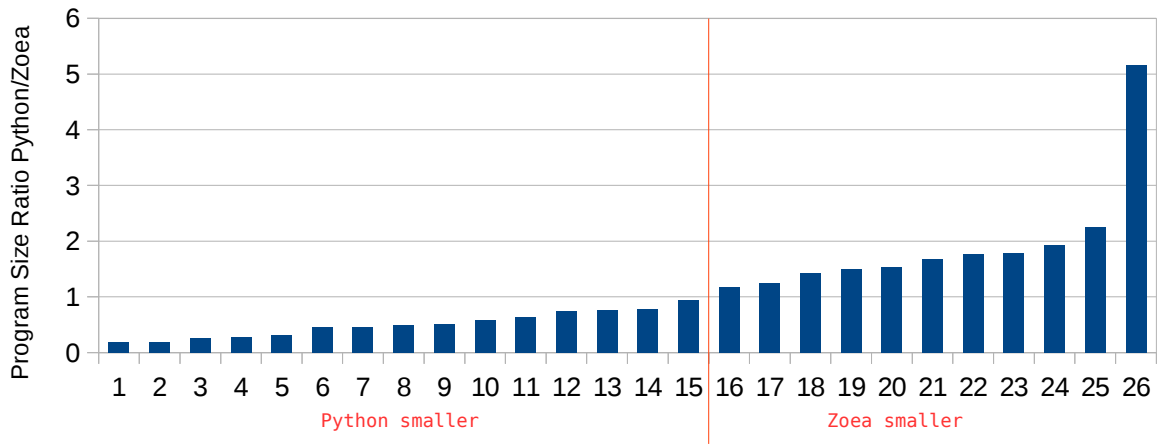

**Figure 7 – Program Size Ratio Python/Zoea Ranked By Task**

In these plots ratio values greater than 1 indicate cases where Zoea code is less complex or smaller than Python. On the other hand ratio values less than 1 indicate cases where the Python code was less complex or smaller than Zoea. In three cases the cyclomatic complexity of both Python and Zoea programs for the same task is 1 meaning that complexity is the same and the ratio is also 1.

For cyclomatic complexity 19 of the tasks (70%) see a drop in complexity from Python to Zoea. Three of the tasks (11.5%) have the same cyclomatic complexity for both Python and Zoea and five tasks (19%) have higher cyclomatic complexity with Zoea as opposed to Python.

With Halstead difficulty 23 tasks (88.5%) have lower complexity with Zoea than with Python whereas only three tasks (11.5%) have higher complexity with Zoea.

In terms of program size 15 tasks (57.6%) have larger programs in Zoea while the remainder are smaller. However the average size of Zoea programs across all tasks (123.88 bytes) is within 1% of the average size of Python programs across all tasks (126.34 bytes). Similarly the total size of Zoea programs for all tasks (3221 bytes) is within 2% of the total size of Python programs for all tasks (3285 bytes). It can be concluded that on average Zoea code is the same size as equivalent Python code but less complex.

The following are the median values for the ratios between Python and Zoea for all tasks:

- Median cyclomatic complexity ratio: 2.00
- Median Halstead difficulty ratio: 1.93

These are summarised in Figure 8.

In general Halstead difficulty seems to be a more appropriate metric for measuring the complexity of Zoea programs. This is partly due to the concept of operands aligning well with the inductive programming approach. Also as Zoea lacks explicit notations for control structures and conditional logic these features are not so apparent in the test cases.

## 6. Discussion

It is not a given that the use of a simpler programming language will result in simpler solutions and indeed it might be anticipated that the opposite behaviour should be observed. A less complex language might be expected to be less expressive so the representation of a given concept would be expected to be larger and perhaps also involve more complex constructs.

The distinction between essential versus accidental complexity [8] means that at least some of the complexity of any problem is irreducible. The production of simpler solutions using a simpler programming language therefore represents something of a mystery.

The only plausible explanation is that the difference in complexity represents additional accidental complexity that is associated with the use of conventional programming languages. This implies that the essential complexity of the problems included in the study is lower than 50% of the observed software complexity. This is at odds with Brooks [8] view that accidental complexity is largely a solved problem. It also implies that it is currently only through the use of composable inductive programming that this reduction in complexity can be observed. If this is the case then the implications for software development across the board would be profound. For example half the effort currently expended in software development is associated with the complexity of the programming language rather than the complexity of delivering the requirements.

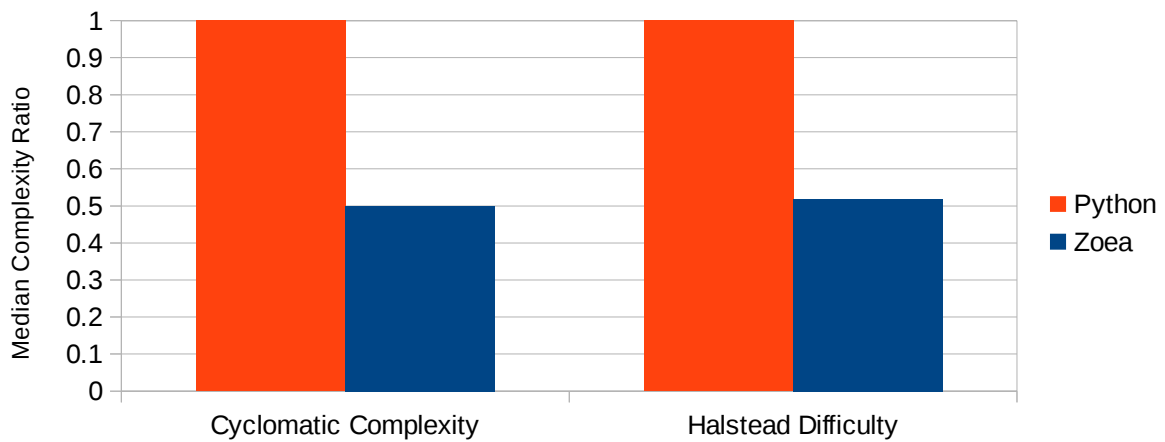

**Figure 8 – Median Complexity Ratios Python/Zoea**

The results could also be interpreted as providing upper bounds for the essential complexity of each task. This is because the complexity of each Zoea program includes both the essential complexity for the relevant task and also it is assumed some accidental complexity arising from the use of the Zoea language. It is currently unknown how any complexity value breaks down into essential and accidental components. Assuming the essential complexity for each task remains constant across different programming languages it is clear that the accidental complexity for Zoea solutions is significantly lower than that for conventional languages. Given the difference in complexity is approximately 50% of the higher value and assuming that Zoea has some tangible accidental complexity then it can be concluded that accidental complexity is the largest single component in current measures of software complexity.

The results of the current study with respect to program size are also a little surprising. It might be expected for Zoea programs to be larger than equivalent programs in conventional languages. This is because all program behaviour in Zoea must be described in terms of data which is neither concise or expressive. Zoea programs are quite wordy particularly when compared to a concise language such as Python. This is because the Zoea language was originally optimised for human readability and simplicity rather than brevity. It is possible that the lack of a significant increase in size is in some way related to the reduction in accidental complexity discussed above. Further investigation would be required to explore this possibility.

Baniassad and Myers [15] introduce the idea that the code that constitutes a program actually forms a higher-level, program specific language which they call 'program language'. In their scheme translation of a program from one programming language to another involves translation between two natural program languages. In this model the complexity of any problem is split between that which lies in the program language and the programming language. The current results appear to support that conjecture.

It is interesting to take this idea a little further. For any given problem a program in a given programming language is a manifestation of the corresponding software requirements. Since most conventional languages are fundamentally very similar it is not difficult to map between actual requirements and elements of the software solution. Such a mapping is not so obvious with Zoea although clearly the requirements must also exist in some form in an inductive programming specification. Analysis of this sort might shed some light on the apparent reduction in complexity observed here. It would also be interesting to understand the nature of the higher level commonality that must exist between conventional programming languages and inductive ones.

For any given task there must be a minimum value in terms of the software complexity required to specify or implement that task. This is distinct from the Komologorov complexity [16] which is a measure of the smallest program to produce a given output. Such a metric would represent a significant part or perhaps all of the essential complexity of a task. It would be useful to be able to quantify or even extrapolate this value.

There are a number of other ways in which the current study could be extended. The study includes a fairly small number of programs and none of these programs could be characterised as large. While the results are adequate to demonstrate a reduction in complexity they raise a number of interesting questions. It would be useful to repeat the study with a larger number of problems, and with problems that are significantly larger in size and complexity. It would also be useful to include other conventional programming languages for comparison.

Beyond the study it would also be interesting to compare the relative software complexity exclusively between different conventional programming languages. We might expect to see significant variability in the software complexity of solutions in different conventional languages – perhaps with some consistency in the complexity ratios for given pairs of languages.

## 7. Conclusions

This paper has presented the results of a study to investigate the impact of composable inductive programming on software complexity. It has compared the complexity and size of a number of programs in the inductive programming language

Zoea with equivalent programs in the popular conventional programming language Python. It has shown that Zoea programs have approximately half the complexity of Python programs according to two established software complexity metrics. Zoea programs are also approximately equal in size to corresponding Python programs. The best explanation for these results is that current programming languages rather than particular software requirements are the source of most of the software complexity encountered by developers. Inductive programming with languages like Zoea presents an opportunity to avoid the significant impact of much of this complexity.

## Acknowledgements

This work was supported entirely by Zoea Ltd. (https://www.zoea.co.uk) Zoea is a trademark of Zoea Ltd. All other trademarks mentioned in this paper are the property of their respective owners.

The authors are grateful to the developers of multimetric [14] which was used in this study under the BSD License.

## References

[1] Rich, C., Waters, R.C. (1993) Approaches to automatic programming. Advances in Computers 37, 1-57. Boston, Academic Press.

[2] Manna, Z., Waldinger, R. (1980) A deductive approach to program synthesis. ACM Transactions on Programming Languages and Systems. 2 (1), 90–121.

[3] Kitzelmann, E. (2010) Inductive programming: A survey of program synthesis techniques. Approaches and Applications of Inductive Programming. Lecture Notes in Computer Science 5812, 50–73. Berlin, Springer-Verlag.

[4] Galwani, S., Hernandez-Orallo, J., Kitzelmann, E., Muggleton, S.H., Schmid, U., Zorn, B. (2015). Inductive Programming Meets the Real World. Communications of the ACM 58 (11): 90–99.

[5] McDaid, E., McDaid, S. (2019) Zoea – Composable Inductive Programming Without Limits. arXiv:1911.08286 [cs.PL].

[6] McCabe, T. (1976). A Complexity Measure. IEEE Transactions on Software Engineering (4): 308–320.

[7] Halstead, M. (1977). Elements of Software Science. Amsterdam: Elsevier North-Holland, Inc. ISBN 0-444-00205-7.

[8] Brooks, F. (1987). No Silver Bullet – Essence and Accident in Software Engineering. IEEE Computer (4): 10–19.

[9] MacLennan, B. (1984) Simple metrics for programming languages. Information Processing & Management, 20 (1–2): 209-221.

[10] Singer, J., Cameron, C., Alexander, M. (2014) Programming Language Feature Agglomeration. In: Workshop on Programming Language Evolution 2014 (PLE14), Uppsala, Sweden, 28 Jul 2014, 11-15. ISBN 9781450328876.

[11] Vanderwiel, S., Nathanson. D., Lilja, D. (1998) A comparative analysis of parallel programming language complexity and performance. Concurrency and Computation Practise and Experience 10 (10): 807-820.

[12] Rosetta Code. Available from: https://rosettacode.org/ (Retrieved: 13 May 2020).

[13] Nanz, S., Furia, C. (2015) A Comparative Study of Programming Languages in Rosetta Code. Proceedings of the 37th International Conference on Software Engineering (ICSE'15): 778-788.

[14] Multimetric 1.1.4. Available from: https://pypi.org/project/multimetric/ (Retrieved: 13 May 2020).

[15] Baniassad, E., Myers, C. (2009) An exploration of program as language. ACM SIGPLAN Notices 44(10): 547-556.

[16] Kolmogorov, A. (1965) Three Approaches to the Quantitative Definition of Information. Problems Inform. Transmission. 1 (1): 1–7.